\documentclass[12pt,a4paper]{article}
\usepackage[latin1]{inputenc}
\usepackage[english]{babel}
\usepackage{amsmath}
\usepackage{amsfonts}
\usepackage{textcomp}
\usepackage{amssymb}
\usepackage{physics}
\usepackage{tensor}
\usepackage{float}
\usepackage{times}
\usepackage{enumerate} 
\usepackage[nottoc,notlot,notlof]{tocbibind}
\usepackage{graphicx}
\usepackage[latin1]{inputenc}
\usepackage[usenames]{color}
\usepackage[left=3.5cm,right=2.5cm,top=2.5cm,bottom=2.5cm]{geometry}
\usepackage{color}
\usepackage{latexsym}
\usepackage{amsmath}
\usepackage{amsfonts}
\usepackage{amssymb}
\usepackage{indentfirst} 
\usepackage{tikz}

\title{Fermat's principle in constant gravitational field}
\author{Joanna Gonera\thanks{e-mail: joanna.gonera@uni.lodz.pl},\quad  Piotr Kosi\'nski\thanks{e-mail: piotr.kosinski@uni.lodz.pl}, \quad  Joanna Piwnik\thanks{e-mail: joanna.piwnik@edu.uni.lodz.pl}\\\\
\small Faculty of Physics and Applied Informatics, \\
\small University of \L\'od\'z,\\
\small Pomorska 149/153, 90-236 {\L}\'od\'z, Poland
}
\date{}
\begin{document}
\maketitle 

\begin{abstract}
Recently (Int.Journ.Mod.Phys. {\bf D27} (2018), 1847025) an interesting property of closed light rings in Kerr black holes has been noticed. We explain its origin and derive a slightly more general result.
\end{abstract}

\section*{}

In the recent paper \cite{Hod} it has been shown that the unique null circular geodesics corresponding to closed light ring provides the fastest way  to circle around  Kerr black hole, as measured by asymptotic observers. 
Below we derive this (and slightly more general) result from general Fermat's principle in constant gravitational field.

\begin{figure}
\begin{center}
\begin{tikzpicture}[smooth cycle]
\draw plot[tension=0.5] coordinates{
(0cm,0cm)  
(3cm,0.5cm)  
(6cm,1cm) 
(7cm,3cm)
(8cm,5cm)
(4cm, 5cm)};
\draw (6cm,1cm) node[anchor=north west]{$C'$};
\draw plot[tension=0.5] coordinates{
(1cm, 0.5cm)
(2.5cm, 0.5cm)
(5.1cm, 1.5cm)
(6cm, 1.5cm)
(7cm, 4cm)
(4cm, 4.5cm)
};
\draw (7cm, 4cm) node[anchor=north east]{$C$};
\end{tikzpicture}
\end{center}
\caption{Closed light curve $C$ and its variation $C'$}
\label{CurveFigs}
\end{figure}
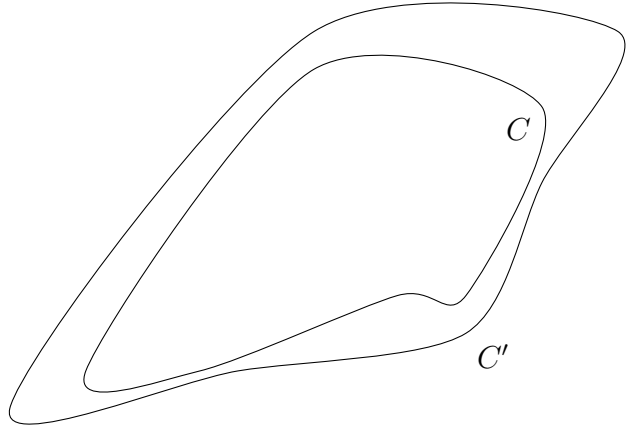

Assume we are dealing with constant gravitational field, i.e. we can choose a system of reference in which all components of the metric tensor are independent of time coordinate $x^0$ called then the world time. It is well known \cite{Landau} that the geometry of light rays is described by the Fermat principle of the form

\begin{equation}
\label{e1}
\delta \int \left( \frac{dl}{\sqrt{g_{00}}} - \frac{g_{0i}dx^i}{g_{00}} \right) = 0
\end{equation}
where

\begin{equation}
\label{e2}
dl^2 = \left( -g_{ij} + \frac{g_{0i}g_{0j}}{g_{00}} \right)dx^idx^j
\end{equation}
defines the metric in three-dimensional space (we adopt the convention $g_{ \mu \nu }=$ diag$(+---)$).\\
Let us note the following relation \cite{Landau} 

\begin{equation}
\label{3}
ds^2=g_{00}\left( dx^0 + \frac{g_{0i}dx^i}{g_{00}}\right)^2 -dl^2
\end{equation}
In eq.(\ref{e1}) one varies over the paths $x^i(\lambda )$ ($\lambda$ being an arbitrary parameter) with fixed ends. Once such a stationary path is determined one recovers null geodesic in space-time by solving $ds^2=0$ with respect to $dx^0$. It is then easily seen that the Fermat principle (\ref{e1}) may be rewritten as

\begin{equation}
\label{e4}
\delta \int dx^0=0
\end{equation}

i.e. null geodesics are stationary with respect to the world time.\\
Consider now a closed light curve, i.e. a closed path $C$ in three-dimensional space, $x^i =x^i( \lambda )$, $\lambda_{0} \leq \lambda \leq \lambda_1$, $x^i(\lambda_0)=x^i(\lambda_1)$, obeying (\ref{e1}). Let $C'$, $x^i=x^i(\lambda ) + \delta x^i(\lambda)$, $\delta x^i(\lambda_0) = \delta x^i(\lambda_1)$, be some neighbouring closed path (not necessarily the light ray).
This situation is presented in Figure~\ref{CurveFigs}.

The integral (\ref{e1}) should be stationary with respect to variations $\delta x^i(\lambda)$. However, in order this to be the case one has to be allowed to neglect the boundary terms. This for sure holds true if we consider the paths with fixed ends. On the other hand, for closed path the boundary terms cancel due to the periodicity. Therefore, the stationarity is preserved also if we consider neighbouring closed path.\\
Now, by solving $ds^2=0$ one can extend both $C$ and $C'$ to the null closed curves in space-time; $C$ becomes a null geodesic while $C'$ is, in general, only null closed curve. From eq.(\ref{e4}) we conclude that the word time is stationary with respect to the variations $\delta x^i(\lambda)$.

Consider now the particular case of Kerr black hole. According to the above reasoning the null circular geodesic may be described as providing the stationary time, as measured by asymptotic observers, to circle around Kerr black hole. However, the reverse is also true. Assume a given circle (not necessarily the geodesic) provides the stationary time to circle around the Kerr black hole with the velocity of light. One can rewrite eq.(\ref{e4}) in the form

\begin{equation}
\label{e5}
\delta \int_{0}^{2\pi}\frac{dx^0}{d\phi}d\phi =0
\end{equation}

while solving $ds^2 = 0$ yields

\begin{equation}
\label{e6}
\frac{dx^0}{d\phi}= F\left( r,\sin^2{\theta},\left(\frac{d\theta}{d\phi}\right)^2, \left( \frac{dr}{d\phi}\right)^2\right)
\end{equation}

Therefore, the variational derivative of the integrand in eq.(\ref{e5}) with respect to $\theta$ vanishes identically for $\theta \equiv \frac{\pi}{2}$ while the one with respect to $r$ is constant on any circle $r=$ const. It follows than that in order to verify the validity of eq.(\ref{e5}) it is sufficient to consider the variations $\delta r=$ const.\\
Concluding, there is one-to-one correspondence between circular null geodesics around Kerr black hole and the circular null geodesics around Kerr black hole and the circular null path corresponding to the stationary world time. Noting that there is the unique null circular geodesic in Kerr metric one concludes that it provides the fastest way to circle around black hole.\\
Actually, this reasoning is valid for any axially symmetric metrics obeying eq.(\ref{e6}) with $x^0$ being the world time and global extremum replaced by local one(s).

\end{document}